\documentclass[a4paper,aps,pra,showpacs,preprintnumbers]{revtex4}
 \usepackage[cp1251]{inputenc}
 \usepackage{amsmath}
 \usepackage{graphics,graphicx,epsfig}
\usepackage{verbatim}

\begin{document}

\def\BE{\begin{equation}}
\def\EE{\end{equation}}
\def\BY{\begin{eqnarray}}
\def\EY{\end{eqnarray}}
\def\L{\label}
\def\nn{\nonumber}
\def\({\left (}
\def\){\right)}
\def\[{\left [}
\def\]{\right]}
\def\<{\langle}
\def\>{\rangle}
\def\o{\overline}
\def\BA{\begin{array}}
\def\EA{\end{array}}
\def\dsp{\displaystyle}
\def\ds{\displaystyle}
\def\k{\kappa}
\def\dd{\Delta}

\title{
Quantum correlated pulses from a synchronously pumped optical
parametric oscillator operating above threshold}

\author{ Valentin A. Averchenko, Yuri M. Golubev }
\affiliation{V.~A.~Fock Physics Institute, St.~Petersburg State University \\
198504 St.~Petersburg, Stary Petershof, ul. Ul'yanovskaya, 1,
Russia}
\author{Claude Fabre, Nicolas Treps}
\affiliation{Laboratoire Kastler Brossel, Universit\'{e} Pierre et
Marie Curie-Paris6,
 Place Jussieu, CC74, 75252 Paris
Cedex 05, France}
\date{\today}

\begin{abstract}

We perform the quantum analysis of the light emitted by a synchronously pumped
optical parametric oscillator operating in the above threshold
regime, i.e. when the peak power of pulsed pumping exceeds the threshold of
continuous generation.
We show that both regimes (below and above threshold) are realized
at different times within each pulse of the signal field.
We show that the quantum fluctuations of signal and pump pulses are
correlated between nearby pulses at times that are placed in the
same position relative to the center of the pulses, whereas
fluctuations are totally not correlated at different times within
the same pulse.
The model also predicts the existence of cross-correlations between
pump and signal pulses.

It is shown theoretically that there is suppression of quantum noise
at the frequencies multiple of the pulse repetition frequency in the
spectra of phase quadratures of pump and signal fields measured by a
balanced homodyne detection with a pulsed local oscillator.
\end{abstract}
\maketitle

\section{Introduction}

Development and investigation of effective sources of non-classical
multimode light are among current trends of quantum optics
\cite{Wasilewski:06, Lassen:07, Chalopin:09}.
These sources appear as a main component required for parallel
quantum information protocols \cite{Leuchs:book:07, Menicucci:08}.
From this point of view an optical parametric oscillator (OPO)
pumped by a mode-locked laser with a period of pulses which is equal
to the round-trip time of the oscillator cavity (so called
synchronously pumped OPO or SPOPO \cite{Driel:spopo:95}) looks quite
promising.
A modal approach used in \cite{Valcarce2006, PateraG.2010} showed
that quantum state of signal light of a SPOPO operating below
threshold in a degenerate regime generates a tensor product of
squeezed states in different super-modes, each being a particular
coherent superposition of longitudinal modes of different
frequencies.
Therefore the signal light generated by a SPOPO appears as an interesting resource for
parallel quantum information protocols such as quantum
teleportation and quantum key distribution.

The same system was also considered in the time domain
\cite{Averchenko2010} using a simple physical model that neglects
all dispersion effects in the parametric crystal. It was shown in
particular that the quantum fluctuations of signal field are not
correlated at different times within each individual pulse whereas
they are correlated between nearby pulses at times that are placed
in the same position relative to the center of the pulses. This
paper dealt with the SPOPO operating below threshold. However it is
well known that the light of a continuously pumped OPO also reveals
nonclassical properties above the oscillation threshold.
\cite{Walls:book:95}.
Therefore the question about the quantum properties of SPOPO light above
threshold looks quite natural.

In the paper \cite{Chalopin:through_threshold:10} it was predicted
that a multimode OPO (i.e. which is simultaneously resonant for
several spatial or temporal modes) above threshold generates bright
light in a single mode whereas the other modes stay in noncritically
squeezed vacuum states, i.e. the amount of squeezing does not depend
on threshold excess.
The prediction was confirmed by an experiment on parametric
generation of TEM$_{10}$ and TEM$_{01}$ modes in squeezed vacuum
states when the threshold is crossed.
In our case it is interesting to check this prediction for the SPOPO,
which is another multimode system.

In this paper we will analyze the above threshold behaviour of SPOPO in
the time domain where the system operation is described by two coupled
equations for envelopes of pump and signal pulses.
%
%The approach looks quite natural for description of pulsed system.
%
The physical model and the two-time technique used to describe the
field evolution inside the oscillator cavity are similar to the ones
applied in \cite{Averchenko2010} for the analysis of the below
threshold regime.

The paper is organized as follows.
In Sec. \ref{II} we present our model of SPOPO and write
Heisenberg-Langevin equations that describe in time domain its
operation above the oscillation threshold.
In Sec. \ref{III} we estimate classical steady-state envelopes of
generated pump and signal pulses. We assume that the pump pulses of
oscillator have a rectangular shape.
We determine quantum fluctuations of fields around the steady-state
values in Sec. \ref{sec_qnt_fl}. For this purpose we solve
linearized Heisenberg-Langevin equations applying adiabatic
elimination of the pump field.
As a result we calculate and analyze pair correlation functions for
the quadrature components of the output pulses.
In the Sec. \ref{sect_BHD} we consider the model of balanced
homodyne detection of pulsed field and calculate the spectra of the
photocurrent fluctuations.
We generalize the results obtained for rectangular pump and local
oscillator pulses to the case of arbitrary envelopes of pulses.

\section{Physical model and main equations}\L{II}

A quantum analysis of the SPOPO below threshold on the basis of time
approach has been performed in \cite{Averchenko2010}.
As a result two-time correlation functions were obtained  for the
output signal field which correspond to the following picture of the
system operation. Under pulsed laser pumping of a $\chi^{(2)}$
nonlinear crystal (see Fig. {\ref{fig:SPOPO}}), placed inside a ring
optical cavity of the oscillator, pump photons are parametrically
down-converted into pairs of correlated signal photons.
The photons of each pair may leave the cavity in different pulses,
during an overall time of the order of photon lifetime in a cavity
$\kappa_s^{-1}$, giving rise to temporal correlations on the same
range of time difference. Approximations of thin nonlinear crystal
and instantaneous down-conversion, used in the work, lead to the
peculiarity that field fluctuations at different times in a single pulse are not correlated.

We will consider SPOPO operation above oscillation threshold on the
basis of the physical model that we have used in the below threshold situation.
%
%We will consider SPOPO operation above oscillation threshold on the
%basis of physical model, used for analysis of below threshold
%regime.
%
Similarly we assume a degenerate parametric interaction for
the carrier frequencies of the pump and signal modes, meaning that
the following phase matching condition is fulfilled $\Delta  k =
k_p(\omega_{p})-2 k_s(\omega_{p}/2) =0$, where $k_p(\omega_{p})$ and
$k_s(\omega_{p}/2)$ are wave vectors of carriers of pump and signal
fields.
The SPOPO ring cavity is resonant and of high-finesse both for pump and
signal fields, so that the doubly-resonant configuration is
realized. It takes the same time $T_R$ for pump and signal pulses to
make a single round trip inside the cavity and this time is equal to
the pump repetition rate.
We assume also that the cavity is dispersion-compensated by
intracavity dispersive elements. This implies that optical pulses of
arbitrary shapes are not distorted after one round trip inside the
cavity.

The above threshold regime is achieved by increasing the pump power.
In this regime pump depletion becomes significant as well as the
process of up-conversion of pairs of signal photons into pump
photons.
Under these conditions the system becomes nonlinear. In particular one can
expect that correlations between the pump and signal pulses will develop.
Therefore an adequate system description must consider the coupled
evolutions of both fields.

In order to describe system operation we use the time-domain approach
presented in detail in Appendix~\ref{A}.
The Heisenberg-Langevin equations obtained under this approach have
the following form \cite{Averchenko2010}
    \begin{eqnarray}
    && \frac{\partial \hat{A}_p(t,T)}{\partial T} = - \kappa_p
    \(\hat{A}_p(t,T) -{A}_0(t)\)
    - g {\hat{A}_s^{\; 2}(t,T)} + \hat{F}_{p}(t,T), \L{HL_p}\\
    && \frac{\partial \hat{A}_s(t,T)}{\partial T} = - \kappa_s
    \hat{A_s}(t,T) + 2g { \hat{A}_p(t,T) } \hat{A}_s^{\dag}(t,T) +
    \hat{F}_{s}(t,T) \L{HL_s}
    \end{eqnarray}
Here $\hat A_p(t,T)$ and $\hat A_s(t,T)$ are quantized envelopes of
pump and signal pulses inside the cavity. Time argument $t$ is
treated as the time deviation from the pulse center and it changes
in the interval from $-T_R/2$ to $T_R/2$; the dependence of
envelopes on the second time argument $T$ describes their changes
from pulse to pulse.
The amplitudes are normalized so that values
$\langle\hat{A}_{p,s}^{\dag} \hat{A}_{p,s}\rangle$ have the meaning
of mean fluxes in photons per second through the cross sections of
the light beams.
$\kappa_{p}$ and $\kappa_{s}$ are the loss rates of pump and
signal fields respectively; $g$ is a constant characterizing the
parametric coupling; ${A}_0(t)$ is the classical steady-state
envelope of the pump pulses inside the cavity, depending only on
time $t$ since the pump pulses are supposed to be perfectly identical.
One can write this envelope in the following general form
    \begin{eqnarray}
    && A_0(t)= \sqrt{N_0(t)} \; e^{i \phi_{in}(t)},
    \end{eqnarray}
where $N_0(t)$ is the intensity shape of the pump pulses and $\phi_{in}(t)$  their possible phase modulation.

The last terms in equations $\hat{F}_{p}$ and $\hat{F}_{s}$ describe
Langevin noise sources which are related to the vacuum fluctuations of the
incoming fields.
In this case the fields have zero mean values and characterized by the
following nonzero pair correlation functions (for details see
Appendix~\ref{B})
    \begin{eqnarray}
    && \langle \hat{F}_{r}(t,T) \hat{F}_{r}^{ \dag}(t',T') \rangle =
    2\kappa_r \; \delta(T-T')
    \delta(t-t'), \qquad r=s,p \L{noise}
    \end{eqnarray}

\section{Oscillation threshold and semiclassical steady-state
solutions}\L{III}
As it is known (see, for example, \cite{Walls:book:95}), an optical
parametric oscillator is a system that exhibits properties of a
second-order phase transition when pump power achieves a certain
threshold value. In our notation the threshold power reads
    \begin{eqnarray}
    && N_{th} \equiv \frac{\kappa_s^2}{4g^2}. \L{N_th}
    \end{eqnarray}
For pulsed pumping the threshold is achieved when the peak power of
pulses is equal to this value. Evidently, corresponding mean power
can be much less than the threshold value. This is an important advantage of pulsed pumping with respect to the continuous one.

It is important to note that in the Heisenberg-Langevin equations
(\ref{HL_p}) and (\ref{HL_s}) time $t$ appears as a parameter and
the Langevin noise sources are delta-correlated relative to this
time according to (\ref{noise}). Hence in our model different
temporal parts of an individual pulse (pump or signal one) develop
in time $T$ independently of other parts with corresponding
instantaneous value of the external pump power $N_0(t)$. As a result
both regimes (below and above threshold) could be realized within
the single pulse at different time instants if peak pump power
exceeds threshold value (see Fig. {\ref{fig:threshold}}). This
situation is analogous to what happens in the spatial domain in a
degenerate confocal cavity\cite{Marte:patterns:98}. Since the
properties of pulsed fields below threshold were analyzed in
\cite{Averchenko2010} we will consider further the simplest case,
when external pump pulses have rectangular temporal shape of
duration $\tau_p$ with intensity exceeding threshold one.
%In this case it becomes possible
%to analyze pulsed fields generated above threshold.
In the last section we will generalize our results to arbitrary envelopes
of the pulses.

Let us define a time dependent pump parameter:
    \begin{eqnarray}
    && \mu(t) \equiv \sqrt{\frac{N_0(t)}{N_{th}}}
    \end{eqnarray}
For rectangular pump pulses the parameter is non-zero $\mu_0
> 1$ when $t$ changes from $-\tau_p/2$ to $\tau_p/2$.
%
%Under these conditions generated signal and pump fields with
%envelopes $\hat{A}_s(t,T)$ and $\hat{A}_p(t,T)$, respectively, have
%the following properties.
%
Between pump pulses the generated fields are vacuum noise while
within the pulses they develop in a way that is
described by coupled Heisenberg-Langevin equations. At this time
intervals the fields are characterized by non zero average amplitudes.
Here we
determine the average steady-state amplitudes denoted as ${A}_p$ and
${A}_s$. In the next section we will analyze the properties of the fluctuations around the average
The amplitudes must satisfy equations which are the classical
counterparts of Heisenberg-Langevin equations (\ref{HL_p}) and
(\ref{HL_s}).
Technically, we replace the operator quantities in these
equations by the c-number quantities and to reject the noise
sources. Finally setting partial derivatives equal to zero, we find
the stationary solutions in the form
    \begin{eqnarray}
    && {A}_p(t) = \sqrt{N_p}\;e^{i \phi_{in}(t)},   \qquad
    {A}_s(t) = \pm\sqrt{N_s}\;e^{i \phi_{in}(t)/2} \L{A_st}
    \end{eqnarray}
Classical intensities $N_p$ and $N_s$ are defined by the following
expressions
    \begin{eqnarray}
    && N_p = N_{th}, \quad N_s = {\frac{2\kappa_p}{
    \kappa_s} \left(\mu_0 - 1 \right)}N_{th}, \L{N_st}
    \end{eqnarray}
which agree with the fact that the average amplitude of pumping
field inside the cavity (and respectively at the output) does not
depend on the pump intensity\cite{Zernike:book:06}. Also one can see
that there are two possible opposite phases of the signal field for
a fixed pump phase.
One can show that both solutions are stable \cite{Walls:book:95}.

\section{Solutions of linearized Heisenberg-Langevin equations} \L{sec_qnt_fl}
The following expressions define the quantum fluctuations of envelopes of
pump $\delta \hat{A}_p$ and signal $\delta \hat{A}_s$ fields within
the pulses (i.e. at time interval $ -\tau_p/2 \leq t \leq \tau_p/2$)
    \begin{eqnarray}
    && \hat{A}_p(t, T) = (\sqrt{{N}_p} + \delta \hat{A}_p(t, T))e^{i
    \phi_{in}(t)},
    %\qquad |t| \leq \tau_p/2
    \L{dA_p} \\
    && \hat{A}_s(t, T) = (\pm\sqrt{{N}_s} + \delta \hat{A}_s(t, T))e^{i
    \phi_{in}(t)/2} \L{dA_s}
    \end{eqnarray}
Let us assume that the fluctuations are small compared to the mean
values
    \begin{eqnarray}
    && \delta \hat{A}_r(t, T) \ll \sqrt{N_r}, \qquad r=p,s \L{flct_cnd}
    \end{eqnarray}
It will be shown later that above threshold these inequalities are
well-satisfied.
Substituting expressions (\ref{dA_p}) and (\ref{dA_s}) into
(\ref{HL_p}) and (\ref{HL_s}) and ignoring the second-order terms
according to the assumption of small fluctuations, one obtains
the following linearized equations for the fluctuations of fields
    \begin{eqnarray}
    && \frac{\partial }{\partial T} \delta\hat{A}_p(t,T)=
    - \kappa_p \delta\hat{A}_p(t,T) \mp 2g \sqrt{{N}_s}\;\delta \hat{A_s}(t,T)+ \hat{F}_{p}(t,T), \L{HL_p_lin}\\
    && \frac{\partial}{\partial T}  \delta\hat{A_s}(t,T) =
    - \kappa_s \delta\hat{A_s}(t,T) + 2g \sqrt{ {N}_p } \;
    \delta\hat{A}_s^{\dag}(t,T) \pm
    2g \sqrt{ {N}_s } \; \delta\hat{A}_p(t,T) + \hat{F}_{s}(t,T) \L{HL_s_lin}
    \end{eqnarray}
These equations can be conveniently solved by writing them in terms of
quadrature components of fields defined as real and imaginary parts
of complex amplitudes:
    \begin{eqnarray}
    && \delta\hat{A}_{r}(t,T) = \delta\hat X_{r}(t,T) + i \delta\hat
    Y_{r}(t,T), \L{quadr_def}
    \end{eqnarray}
where $\delta\hat{X}_r = \delta\hat{X}_r^\dag$ and $\delta\hat{Y}_r
= \delta\hat{Y}_r^\dag$. This choice of quadrature components is such
that their fluctuations define fluctuations of intensities and
phases of fields, respectively.
These quadrature components can be experimentally measured using balanced homodyne detection of fields under appropriate choice of
the local oscillator field that will be considered in the Sec.
\ref{sect_BHD}.

One gets the following equations for quadrature components of pumping
field
    \begin{align}
    \frac{\partial }{\partial T} \delta\hat{X}_p(t,T) &=
    - \kappa_p \delta\hat{X}_p(t,T) \mp 2g \sqrt{N_s} \; \delta \hat{X_s}(t,T)+ \hat{F}_{p}'(t,T), \L{HL_xp}\\
    \frac{\partial }{\partial T} \delta\hat{Y}_p(t,T) &=
    - \kappa_p \delta\hat{Y}_p(t,T) \mp 2g \sqrt{N_s}\;\delta \hat{Y_s}(t,T)+ \hat{F}_{p}''(t,T), \L{HL_yp}
    \end{align}
and signal field
    \begin{align}
    \frac{\partial}{\partial T}  \delta\hat{X_s}(t,T) &=
    %- \kappa_s\delta\hat{X_s}(t,T) + 2g { {A}_p(t) } \delta\hat{X}_s(t,T) +
    \pm 2g \sqrt{N_s} \; \delta\hat{X}_p(t,T) + \hat{F}_{s}'(t,T)
    \L{HL_xs},\\
    \frac{\partial}{\partial T}  \delta\hat{Y_s}(t,T) &=
    - 2\kappa_s\delta\hat{Y_s}(t,T) \pm
    2g \sqrt{N_s}\; \delta\hat{Y}_p(t,T) + \hat{F}_{s}''(t,T) \L{HL_ys}
    \end{align}
Here we took into account that above threshold
$N_p=\kappa_s^2/(4g^2)$ and we also defined hermitian quadrature
components of Langevin sources:
    \begin{align}
    & \hat{F}_{p}(t,T) e^{-i \phi_{in}} = \hat{F}_p'(t,T) + i
    \hat{F}_p''(t,T),\\
    & \hat{F}_{s}(t,T) e^{-i \phi_{in}/2} = \hat{F}_s'(t,T) + i
    \hat{F}_s''(t,T),
    \end{align}
with the following correlation functions, obtained from
(\ref{noise}):
    \begin{align}
    & \langle \hat{F}_{r}'(t,T) \hat{F}_{r}'(t',T') \rangle
    =  \langle \hat{F}_{r}''(t,T) \hat{F}_{r}''(t',T') \rangle
    =  \frac{\kappa_r}{2} \; \delta(T-T') \delta(t-t'),
    \L{noise_quadr}\\
    & \langle \hat{F}_{r}'(t,T) \hat{F}_{r}''(t',T') \rangle
    =  - \langle \hat{F}_{r}''(t,T) \hat{F}_{r}'(t',T') \rangle
    =  i \frac{\kappa_r}{2} \; \delta(T-T') \delta(t-t') \L{Lang_noise_crlt}
    \end{align}
Expression (\ref{Lang_noise_crlt}) shows that the quadratures of fields
are correlated. Thus fluctuations of X- and Y-quadratures of
pump and signal fields will be also correlated.
Nevertheless in the case we are interested in these correlations are
not important.

Let us consider the case when relaxation of the pump field in the
cavity with the rate $\kappa_p$  is the fastest process.
Then equations (\ref{HL_xp})-(\ref{HL_ys}) can be solved using
adiabatic elimination of this field. Setting derivatives equal to
zero in equations (\ref{HL_xp})-(\ref{HL_yp}), the fluctuations
of the pump field have the form
    \BY
    && \left(\begin{array}{c}\delta\hat X_p(t,T)\\\delta\hat Y_p(t,T)\end{array}\right) =
    \mp \sqrt{\frac{\kappa_x}{\kappa_p}} \; \left(\begin{array}{c} \delta\hat{X}_{s}(t,T)\\\delta\hat{Y}_{s}(t,T)\end{array}\right) +
    \frac{1}{\kappa_p} \; \left(\begin{array}{c} \hat{F}_{p}'(t,T)\\\hat{F}_{p}''(t,T)\end{array}\right) \L{p_sol_ad}
    \EY
Inserting these expressions into equations (\ref{HL_xs}) and
(\ref{HL_ys}), one gets simple differential equations which have the
following solutions
    \BY
    && \left(\begin{array}{c}\delta\hat X_s(t,T)\\\delta\hat Y_s(t,T)\end{array}\right)=
    \int_{-\infty}^{T}d{T^\prime} \[\pm \sqrt{\frac{\kappa_x}{\kappa_p}}
    \left(\begin{array}{c}\hat{F}^\prime_{p}(t,
    T^\prime)\\\hat{F}^{\prime\prime}_{p}(t,
    T^\prime)\end{array}\right) + \left(\begin{array}{c} \hat{F}_{s}'(t,T)\\\hat{F}_{s}''(t,T)\end{array}\right) \]
    e^{\dsp -\kappa_{x,y} (T-T^\prime)}, \L{s_sol_ad}
    \EY
where $\kappa_{x} = 2\kappa_{s}(\mu_0-1)$ and $\kappa_{y} =
2\kappa_{s}\mu_0$ are effective damping rates of fluctuations.
According to adiabatic approximation these rates fulfill the
condition $\kappa_{x}, \kappa_{y} \ll \kappa_p$ that limits pump
parameter $\mu_0 \ll \kappa_p/\kappa_s$. We will show further that
the main quantum effects appear close to the oscillation threshold
when $\mu_0\approx 1$, which is in agreement with our restriction.

These solutions enable us to determine the properties of the output
fields which are of practical interest. For this purpose one uses the boundary condition on the output mirror that reads
    \begin{eqnarray}
    && \hat A_r^{out}(t,T)=
    \sqrt{{\cal T}_r} {\hat A}_r(t,T)-\sqrt{1-{\cal T}_r} {\hat A}_r^{in}(t,T). \L{mirror}
    \end{eqnarray}
Here ${\cal T}_r$ is the transmission coefficient of the cavity
mirror related to the loss rate of the field $\kappa_r = {\cal T}_r
/(2 T_R)$ (when ${\cal T}_r \ll 1$).
Also vacuum fluctuations of the incoming field ${\hat A}_r^{in}$ in
the expression are related to Langevin noise sources ${\hat F}_r$
according to (\ref{A-F}).
The boundary condition is also valid for the quadrature components of
fields.

\section{Correlations between pulses}\L{t_crlt}

Let us now determine the correlation functions of the quadratures of
output fields. Before it we turn from time $t$ that describes a
deviation from center of pulses to usual time scale.
%
%Before it we turn from time $t$ describing single pulse to usual
%time scale.
%
For this one returns to the discrete pulse numbering, replacing
time $T$ with discrete number $n$ and time $t$ with $t-nT_R$
in the following way:
    \begin{align}
    \nonumber &T\to nT_R, \quad t\to t-nT_R, \quad T_R\delta(T-T')\to\delta_{nn'},\\
    &\delta\hat X_r^{out}(t,T) \to \delta\hat
    X_{r,n}^{out}(t-nT_R),\quad \delta\hat Y_r^{out}(t,T) \to \delta\hat
    Y_{r,n}^{out}(t-nT_R).
    \end{align}
Thus, using results of the previous section, we obtain the following pair
correlation for the quadrature components of pump pulses
    \begin{align}
    \< \delta \hat X_{p,n}^{out}(t-nT_R) \delta \hat X_{p,n'}^{out}(t'-n'T_R) \>
    &=\frac{1}{4} \(\delta_{nn'}\delta(t-t') + 2\kappa_s T_R e^{-2\kappa_s
T_R(\mu_0-1)|n-n'|}
    \delta(t-t'-(n-n')T_R)\),
    \L{xx_p_n}\\
    \< \delta\hat Y_{p,n}^{out}(t-nT_R) \delta \hat Y_{p,n'}^{out}(t'-n'T_R) \>
    &=\frac{1}{4} \(\delta_{nn'}\delta(t-t') - 2\kappa_s T_R \frac{(\mu_0-1)}{\mu_0} e^{-2\kappa_s T_R \mu_0|n-n'|} \delta(t-t'-(n-n')T_R)\) \L{yy_p_n}
    \end{align}
and signal pulses
    \begin{align}
    \< \delta \hat X_{s,n}^{out}(t-nT_R) \delta \hat X_{s,n'}^{out}(t'-n'T_R) \>
    &=\frac{1}{4} \(\delta_{nn'}\delta(t-t') + \frac{\kappa_s T_R}{\mu_0-1} e^{-2\kappa_s T_R(\mu_0-1)|n-n'|}
    \delta(t-t'-(n-n')T_R)\),
    \L{xx_s_n}\\
    \< \delta\hat Y_{s,n}^{out}(t-nT_R) \delta \hat Y_{s,n'}^{out}(t'-n'T_R) \>
    &=\frac{1}{4} \(\delta_{nn'}\delta(t-t') - \frac{\kappa_s T_R}{\mu_0} e^{-2\kappa_s T_R \mu_0|n-n'|} \delta(t-t'-(n-n')T_R)\). \L{yy_s_n}
    \end{align}
These expressions are similar to the correlation functions of pulsed
signal field generated by SPOPO below oscillation threshold.
The first terms on the right hand side which are proportional to
$\delta(t-t')$ are due to the incoming vacuum field reflected by the
coupling mirror of the cavity. The second terms are related to the
fields coming out of the cavity.
For individual pulses, i.e. when $n=n^\prime$, one can neglect
second terms which are proportional to $\kappa_s T_R \ll 1$. Thus
the fluctuations of the quadratures in the pulses are vacuum fluctuations, except
the fluctuations of X-quadrature which become infinitely large close to
threshold when $\mu_0 \rightarrow 1$.
Since these solutions are obtained under the assumption of small
fluctuations one must define how close can we approach the
threshold from above.
%restrict the minimal value of pump parameter applicable in our
%analysis.
%
It is shown in the Appendix \ref{C} that for typical experimental
parameters the condition of small fluctuations is fulfilled and the
solutions are correct if $(\mu_0-1) \gg 10^{-3}$.
Therefore we will use the following minimal value of the pump parameter in
quantitative estimations $\mu_0=1.1$.

There are also quantum correlations between different pulses (when
$n\neq n^\prime$) for the X- quadrature of the fields, and
anticorrelations for the Y-quadrature with the correlation
coefficient proportional to $\kappa_s T_R$.
The number of significantly correlated successive pulses can be
evaluated by the factor in the exponential, which is proportional to
$(\kappa_s T_R)^{-1}$.
The delta-function shows that the correlations between different
pulses have a ”local” character: they are effective only when the
time differences are a multiple of the period $T_R$.

Correlation coefficient and number of correlated pulses also depend
on pump parameter $\mu_0$ (i.e. on pump power).
For Y-quadrature of the signal field they are maximal close to
threshold, when $\mu_0 \approx 1$.
For Y-quadrature of the pumping field correlation coefficient tends
to zero in the vicinity of threshold. Away from threshold
the correlation increases while the number of correlated pulses decreases.
Also close to threshold the fluctuations of the X-quadrature of signal field
becomes infinitely large, which restricts the minimal value of pump
parameter as remarked above.

Solutions (\ref{p_sol_ad}) and (\ref{s_sol_ad}) also show that
fluctuations of pump and signal fields are correlated in contrast to
below threshold regime. One can get the following symmetrized
cross-correlation functions for these fields at the output of the
oscillator
    \begin{multline}
    \< \delta \hat X_{p,n}^{out}(t-nT_R) \delta \hat X_{s,n'}^{out}(t'-n'T_R)
    \> +
    \< \delta \hat X_{s,n}^{out}(t-nT_R) \delta \hat X_{p,n'}^{out}(t'-n'T_R) \>
    =\\
    - \kappa_s T_R \sqrt{\frac{1}{2(\mu_0-1)}} \; e^{-2\kappa_s T_R(\mu_0-1)|n-n'|} \delta(t-t'-(n-n')T_R),
    \end{multline}

    \begin{multline}
    \< \delta\hat Y_{p,n}^{out}(t-nT_R) \delta \hat Y_{s,n'}^{out}(t'-n'T_R)
    \> +
    \< \delta \hat Y_{s,n}^{out}(t-nT_R) \delta\hat Y_{p,n'}^{out}(t'-n'T_R)  \>
    =\\
    - \kappa_s T_R \sqrt{\frac{(\mu_0-1)}{2\mu^2_0}} e^{-2\kappa_s T_R \mu_0|n-n'|} \delta(t-t'-(n-n')T_R)
    \end{multline}
The properties of the cross-correlations (number of correlated
pulses, correlation coefficient and local character of correlations)
are analogous to the properties of fields itself that we have considered
above. Close to the oscillation threshold the correlations between pump and
signal fields become infinitely large for X-quadrature and decrease
for Y-quadrature. The minus sign indicates that the fluctuations are
anti-correlated.

\section{Quantum effects in the spectra of fields}\L{sect_BHD}

Let us remind that the temporal features associated with the correlations of
pulses are small (of the order of $\kappa_s T_R \ll 1$), while the
number of correlated pulses is defined by inverse value $(\kappa_s
T_R)^{-1}$.
Therefore one expects to obtain important quantum effects by
observing integral characteristic of fields, such as the Fourier spectrum of
fluctuations of quadratures.
Let us consider the measurement of field quadratures obtained by a
balanced homodyne detection of the output fields (see Fig.
{\ref{fig:BHD}).
The signal or pump field is mixed at a symmetric beamsplitter
with an intense local oscillator field of the same optical
frequency ($\omega_p/2$ or $\omega_p$).
In this case the fluctuations of the difference photocurrent are given
by the expression
    \begin{eqnarray}
    &&\delta\hat i_r(t)=\beta^*(t) \; \delta\hat A_r^{out}(t) + \beta(t) \; \delta\hat A_r^{out
    \dag}(t),
    \end{eqnarray}
where $\beta(t)$  is the complex amplitude of the local oscillator.
We suppose here that the local oscillator field is a train of
identical pulses and that their period is equal to the period of analyzed
pulses $T_R$
    \begin{eqnarray}
    && \beta(t)=\sum_{n}\beta_0(t-nT_R).
    \end{eqnarray}
The envelope of pulses has the form
    \begin{eqnarray}
    &&\beta_0(t)= \sqrt{N_{LO}(t)} \; e^{i (\varphi(t) + \Phi)},
    \end{eqnarray}
where phase modulation of pulses is matched with the detected field:
for pump field $\varphi(t) = \phi_{in}(t)$, for signal field
$\varphi(t) = \phi_{in}(t)/2$.
As a result choosing constant phase shift $\Phi=0$ or $\Phi=\pi/2$
one finds:
    \begin{eqnarray}
    &&\delta\hat i_r(t)=2 \sum_n\sqrt{N_{LO}(t_n)}\left(\begin{array}{c} \delta\hat
    X_{r,n}^{out}(t_n)\\\\\delta\hat
    Y_{r,n}^{out}(t_n)\end{array}\right),\qquad t_n=t-nT_R. \L{i_X_Y}
    \end{eqnarray}
The shape of local oscillator pulses, their duration and delay relative
to the pulses of the fields to analyze can be experimentally adjusted.
We suppose here that the pulses have a rectangular shape of duration
$\tau_p$ and are ideally synchronized with the pulses of analyzed field.
In this case the current fluctuations are equal to the quantum
fluctuations of the fields inside pulses.
In Sec. \ref{arb_shape} we will generalize our results to arbitrary envelopes
of pump and local oscillator pulses.

Let us now consider the measurement of Y-quadratures of pump and signal
fields.
Substituting expressions (\ref{yy_p_n}) and (\ref{yy_s_n}) in
(\ref{i_X_Y}), we derive the pair correlation functions for the
currents:
    \begin{align}
    \langle \delta \hat i(t)\;\delta \hat i(t^\prime)\rangle_p=
    &\sum_n N_{LO}(t-nT_R)\(\delta(t-t^\prime) -
    2\kappa_s T_R \frac{(\mu_0-1)}{\mu_0} e^{-2\kappa_s \mu_0|t-t'|}\sum_{n^\prime}
    \delta\(t-t^\prime-(n-n^\prime) T_R\)\), \L{ii_p}\\
    \langle \delta \hat i(t)\;\delta \hat i(t^\prime)\rangle_s=
    &\sum_n N_{LO}(t-nT_R)\(\delta(t-t^\prime) -
    \frac{\kappa_s T_R}{\mu_0} \dsp e^{-2\kappa_s\mu_0|t-t'|} \sum_{n^\prime}
    \delta\(t-t^\prime-(n-n^\prime) T_R\)\). \L{ii_s}
    \end{align}
One sees that the correlations between pulses  lead
to temporal periodic correlations of the photocurrent.
Let us determine the frequency spectrum of the photocurrent quantum
noise defined as:
    \begin{eqnarray}
    && \(\delta i^2\)_\Omega= \lim_{T\to\infty}\frac{1}{T}\int\limits_{-T/2}^{+T/2} dt\int\limits_{-T/2}^{+T/2}
    dt^\prime\;\langle\delta\hat i(t)\;\delta\hat i(t^\prime)\rangle
    \;e^{\dsp i\Omega (t-t^\prime)} \L{iw_def}
    \end{eqnarray}
Substituting (\ref{ii_p}) and (\ref{ii_s}) into (\ref{iw_def}) one
gets the following explicit expressions for the spectrum of the detected pump field
    \begin{eqnarray}
    \frac{(\delta i^2_p)_{\Omega}}{\<I\>}=1 - \sum
    \limits_{m=0,1,2\ldots} \;
    \frac{8\kappa_s^2(\mu_0-1)}{4 \kappa_s^2 \mu_0^2 + \left(\Omega -{
    2\pi m}/{ T_R} \right)^2}, \L{iw_p}
    \end{eqnarray}
and signal field
    \begin{eqnarray}
    \frac{(\delta i^2_s)_{\Omega}}{\<I\>}=1 - \sum
    \limits_{m=0,1,2\ldots} \;
    \frac{4\kappa_{s}^2}{4\kappa_{s}^{2} \mu_0^2 + \left(\Omega -{
    2\pi m}/{ T_R} \right)^2}. \L{iw_s}
    \end{eqnarray}
Here fluctuations are normalized to the corresponding shot (quantum)
noise of local oscillator
    \begin{eqnarray}
    \<I\>=\frac{1}{T_R}\int\limits_{-T_R/2}^{+T_R/2}dt\;N_{LO}(t),
    \end{eqnarray}
which for rectangular pulses is equal to $\<I\> = (\tau_{p}/T_R)
N_{LO}$.

These expressions show that the shot noise is reduced around all
resonant frequencies of the cavity $\Omega_m=2\pi m/T_R$. This coincides with the result obtained in the below threshold regime.
For pump field a maximal noise reduction of fifty percent is
achieved at $\mu_0=2$.
%
%For pumping field shot noise is maximally reduced by half when
%$\mu_0=2$.
%
For signal field the expression predicts full reduction of the noise at
the oscillation threshold, i.e. $(\delta i^2_s)_{\Omega_m}/\<I\>
\rightarrow 0$ when $\mu_0 \rightarrow 1$.

\section{Arbitrary shapes of pulses}\L{arb_shape}

Our results can be easily generalized to arbitrary shapes of
pump pulses with corresponding pump parameter $\mu(t)$. This is
possible because in our model different temporal parts of individual
pulses develop independently of each other inside SPOPO. Hence at
time instants when instantaneous pump power exceeds parametric
threshold, i.e. $\mu(t) > 1$, the average amplitude and fluctuations of
signal field are defined by (\ref{A_st})-(\ref{N_st}) and
(\ref{xx_s_n})-(\ref{yy_s_n}), respectively, where replacement
$\mu_0 \rightarrow \mu(t)$  is made. The average
amplitude of intracavity pump field is stabilized at the threshold value
according to (\ref{A_st})-(\ref{N_st}) and its fluctuations are
characterized by correlations (\ref{xx_p_n}) and (\ref{yy_p_n}). At
other time instants, when $\mu(t) < 1$, the signal field is in the
below threshold regime.

Also it is not difficult to generalize the results of balanced homodyne
detection, namely the spectra (\ref{iw_p}) and (\ref{iw_s}), that we have derived for
rectangular pump and local oscillator pulses. Let us assume that the
local oscillator pulses with the envelope $N_{LO}(t)$ are quite
short ($\tau_{LO}<\tau_{p}$) and synchronized with the analyzed
pulses to probe temporal parts of the pulses generated above threshold. Then
general expressions can be obtained by averaging intensity shape
of the local oscillator pulses with expressions (\ref{iw_p}) and
(\ref{iw_s}), where pump parameter depends on time $\mu(t)$
    \begin{eqnarray}
    && (\delta i^2_p)_{\Omega}= \frac{1}{T_R} \int\limits_{-T_R/2}^{+T_R/2} dt \; N_{LO}(t)
    \(1 - \sum\limits_{m=0,1,2\ldots} \;
    \frac{8\kappa_s^2(\mu(t)-1)}{4 \kappa_s^2 \mu^2(t) + \left(\Omega -{
    2\pi m}/{ T_R} \right)^2}\), \L{iw_pp} \\
    && (\delta i^2_s)_{\Omega}= \frac{1}{T_R} \int\limits_{-T_R/2}^{+T_R/2} dt \; N_{LO}(t)
    \(1 - \sum\limits_{m=0,1,2\ldots} \;
    \frac{4\kappa_{s}^2}{4\kappa_{s}^{2} \mu^2 (t) + \left(\Omega -{
    2\pi m}/{ T_R} \right)^2}\). \L{iw_ss}
    \end{eqnarray}
These expressions mean that current fluctuations at a particular
frequency appear as a weighted sum (integral) of fluctuations from
all non-correlated parts of the measured pulses.
Consequently, the detected quantum noise is sensitive to the
temporal properties of the local oscillator pulses, particularly to
their duration and to the delay relative to analyzed pulses, as in
the below threshold regime.
Fig. {\ref{fig:mode_sqz}} presents suppression of the photocurrent
shot noise at zero frequency as a function of delay $\Delta t$ of
short local oscillator pulses ($\tau_{LO} \ll \tau_p$) relative to
the signal ones.
Curves are obtained for Gaussian pump pulses of SPOPO $\mu(t) =
\mu_0 e^{-2(t/\tau_p)^2}$ and with different values of peak power,
including below threshold values.
In order to determine the noise suppression at the edges of the
signal pulses, where the field is in the below threshold regime, we
used the following expression of the photocurrent spectrum from
\cite{Averchenko2010}
    \begin{eqnarray}
    && \(\delta i^2_s\)_\Omega^{below} = \frac{1}{T_R}\int\limits_{-T_R/2}^{+T_R/2}
    dt N_{LO}(t)
    \(1 - \sum_{m=0, 1, 2,\ldots}\frac{4\kappa_s^2\mu(t)}{\kappa_s^2\(1+\mu(t)\)^2+\(\Omega-2\pi
    m/T_R\)^2}\). \L{i2_w}
    \end{eqnarray}
This expression was also used in the case when the peak pump power is
less than the threshold value, i.e. $\mu_0<1$.

The figure shows that the fluctuations of signal field are nonuniform inside
pulses. In the central part of bright pulses ($\mu_0>1$), where the field
intensity is maximal, the  fluctuations are larger than at the edges.
Of course measuring local squeezing with the help of infinitely
short LO pulses is a purely theoretical scheme.
In the realistic case when pulses of local oscillator have finite
duration, the photocurrent fluctuations must be weighted by the intensity of the
local oscillator pulse that corresponds to expression (\ref{iw_ss}).

\section{Conclusion}

Let us summarize our results:
if the peak pump power exceeds the threshold of continuous OPO oscillation
then both regimes (below and above threshold) coexist at
different times within the pulses of signal field.
Bright parts of the pulses above threshold are
characterized by an average amplitude. At the edges of pulses the field
has properties of the below threshold case.
The quantum fluctuations of pulses turn out to be totally not
correlated at different times within the same pulse, whereas they
are correlated between nearby pulses at times that are placed in the
same position relative to the center of the pulses.
Above threshold the model also predicts the existence of
correlations between pump pulses with the same properties and
cross-correlations between pump and signal pulses.

These correlations can be measured in the scheme of balanced
homodyne detection using a pulsed local oscillator synchronized with
the required times of the analyzed field.
We have also shown that these correlations lead to a suppression of quantum noise
in the spectra of phase quadratures of pump and signal fields around
frequencies $\Omega_m=2\pi m/T_R$ ($m=0,1,2,\ldots$), where $T_R$
period of pulses.
A stronger noise suppression is achieved under detection of signal
field when thelocal oscillator pulses are delayed relative to the peak
of signal ones.

Since a SPOPO is a multimode system it is interesting to compare our
results with the prediction of the paper
\cite{Chalopin:through_threshold:10}. According to this paper, when
one increases the pump power, the multimode parametric oscillator
starts oscillating in the mode with the lowest oscillation threshold
whereas other modes stay in non-critically squeezed vacuum states.

One can consider for SPOPO that the bright parts of the signal pulses
correspond to the modes developing above threshold and the field at
the edges of pulses is formed by the below threshold modes.
The correlation properties of the field show that an individual mode is
a train of correlated delta pulses whereas different modes appear as
trains of pulses delayed to each other.

It is well-known that the balanced homodyne detection technique
detects optical field amplitudes in a particular spatial-temporal
mode defined by a coherent local oscillator pulse \cite{Raymer:91}.
Therefore shot noise suppression at the Fig. {\ref{fig:mode_sqz}}
could be treated as a squeezing of the stated modes of signal field
since different delays of the local oscillator correspond to the
detection of different modes. Further interpretation of the results
in terms of the modes apparently should take into account dispersion
effects in the parametric crystal such as mismatch and dispersion of
group velocities.

The study was performed within the framework of the Russian- French
Cooperation Program ”Lasers and Advanced Optical Information
Technologies” and the European Project HIDEAS (grant No. 221906). It
was also supported by RFBR (No. 08-02-92504).

\bibliography{bible}

\appendix

\section{}\L{A}

In the degenerate parametric generation configuration that we
consider here the field operator inside the cavity is equal to
    \begin{eqnarray}
    && \hat E(z,t)=\hat E_p(z,t)+\hat E_s(z,t).\L{cavity_field}
    \end{eqnarray}
We use the plane wave approximation, so that the field amplitudes
depend only on one longitudinal coordinate $z$ measured along
optical axis of the cavity. As the pump of the parametric crystal is
realized by a train of optical pulses of duration close to $100 fs$
it is possible to disjoint in the standard form quick oscillations
of fields with optical frequencies $\omega_{p,s}$  from slow changes
of their envelopes \cite{Kolobov:RMP:99}.
At the crystal entrance ($z=0$) for the pump $p$ and signal $s$
waves the two fields read
    \begin{eqnarray}
    && \hat E_r(0, t)=i \left(\frac{\hbar \omega_r}{2 n_r \varepsilon_0 c
    S} \right)^{1/2} e^{-i\omega_r t} \; \hat A_r(0,t),\qquad
    r=s,p,\qquad(\omega_p=2\omega_s),\L{slow_amp}
    \end{eqnarray}
where $n_r = n_r(\omega_r)$ are the indices of refraction of the
crystal.
Following expression takes into account the periodic temporal
structure of the fields
    \begin{eqnarray}
    && \hat A_{r}(0,t)=\sum_{n}\hat A_{r,n}(t-nT_R). \L{pulse_def}
    \end{eqnarray}
Here ${\hat A}_{r,n}(t-nT_R)$ is the envelope of the $n$-th pulse.
Argument of the envelope $t-nT_R$ describes time deviation from the
pulse center and changes in the interval from $-T_R/2$ to $T_R/2$.

In order to describe evolution of fields inside the SPOPO cavity we
use the two-time approach applied by Haus in Ref.~\cite{Haus:2004}
for developing a quantum theory of actively mode-locked lasers. We
assume that the envelopes of pulses are not significantly changed
from one pulse to the next, an hypothesis that is typically valid in
experiments with high-finesse cavity and weak parametric
amplification. Then the dependence on discrete number $n$ could be
replaced approximately by the a continuously varying  temporal
parameter $T$ in the following way
    \begin{eqnarray}
    &&\hat A_{r,n}(t-nT_R)\to\hat A_{r}(t,T),\L{T_def}
    \end{eqnarray}
where $t$ on the right hand side of the expression denotes time
deviation from the pulse center.
Thus expressions (\ref{pulse_def}) and (\ref{T_def}) esteblish
correspondence between envelope of the field ${\hat A}_r(0,t)$ and
envelope of an individual pulse ${\hat A}_r(t,T)$ that depends on
two time parameters.

\section{}\L{B}

Let us determine properties of Langevine noise sources entering in
the Heisenberg-Langevine equations (\ref{HL_p}) and (\ref{HL_s}).
The sources are caused by vacuum fluctuations of incoming
coherent/vacuum fields. Following expression directly connects these
quantities
    \begin{eqnarray}
    && \hat{F}_{r}(t) = \frac{\sqrt{{\cal T}_r}}{T_R}\hat A_r^{vac}(t) \approx \sqrt{\frac{2\kappa_r}{T_R}}\hat A_r^{vac}(t), \qquad r=s,p, \L{A-F}
    \end{eqnarray}
where ${\cal T}_r$ is the transmission coefficient of the mirror at
the corresponding frequency. Vacuum fluctuations are characterized
by the following non-zero correlation functions
    \begin{eqnarray}
    && \langle \hat{A}_{r}^{vac}(t) \hat{A}_{r}^{vac \; \dag}(t') \rangle =
    \delta(t-t').
    \end{eqnarray}
Using written expressions and pulse representation of slow envelopes
(\ref{pulse_def}) one gets that noise terms should satisfy following
relations
    \begin{eqnarray}
    && \langle \hat{F}_{r,n}(t-nT_R) \hat{F}_{r,n'}^{\dag}(t'-n'T_R) \rangle =
    \frac{2\kappa_r}{T_R} \delta_{nn'}\delta(t-nT_R-(t'-n'T_R)).
    \end{eqnarray}
Finally making transition to continuous time parameter (\ref{T_def})
and using relations $\delta_{nn'} \to T_R\delta(T-T')$ and $t-nT_R
\to t$, one gets required correlators (\ref{noise}).

\section{}\L{C}

Here we determine values of pump parameter at which the condition of
small fluctuations is fulfilled: $\delta \hat{A}_r(t, T) \ll
\sqrt{N_r}$.
It was shown in Sec.~\ref{t_crlt} that above threshold fluctuations
of fields are vacuum except X-quadrature of signal field whose
fluctuations become infinitely large in the vicinity of threshold
when $\mu_0 \rightarrow 1$. Therefore condition of small
fluctuations could be written in the form
    \begin{eqnarray}
    && \<\delta \hat X_s^2\> \ll N_s, \L{flct_cnd_1}
    \end{eqnarray}
where $\<\delta \hat X_s^2\>$ is a variance of X-quadrature of
intracavity signal field.
In order to determine it let us evaluate correlation function of the
quadrature $\< \delta \hat X_s(t,T) \; \delta \hat X_s(t',T') \>$.
The function could be easily obtained using explicit expression for
fluctuations of X-quadrature (\ref{s_sol_ad}) and correlation
functions for Langevine noise sources (\ref{noise_quadr}). One gets
    \begin{eqnarray}
    \< \delta \hat X_s(t,T) \; \delta \hat X_s(t',T') \>
    = \frac{1}{4} \(1 + \frac{1}{2(\mu_0-1)}\) e^{-2\kappa_s (\mu_0-1)(T-T')} \delta(t-t')
    \end{eqnarray}

We have to put $T=T'$ and $t=t'$  in the expression in order to find
quadrature variance. However divergence appears due to the fact that
field fluctuations are not correlated in each individual pulse.
One can avoid the difficulty defining  dispersion in the following
way
    \begin{eqnarray}
    && \<\delta \hat X_s^2\> = \frac{1}{T_F}
    \int\limits_{t-T_F/2}^{t+T_F/2} dt' \;
    \< \delta \hat X_s(t,T) \; \delta \hat X_s(t',T) \> \L{D_def}
    \end{eqnarray}
where $T_{F}$ is an averaging time.
Thus dispersion is given by
    \begin{eqnarray}
    && \<\delta \hat X_s^2\> = \frac{1}{T_F} \cdot \frac{1}{4} \(1 + \frac{1}{2(\mu_0-1)}\)
    \end{eqnarray}
Finally let us take into account that average classical intensity of
signal field tends to zero in the vicinity of threshold according to
expression $N_s = {\frac{2\kappa_p}{\kappa_s} \left(\mu_0- 1
\right)}N_{th}$. Therefore the condition of small fluctuations is
fulfilled if:
    \begin{eqnarray}
    && \mu_0-1 \gg \sqrt{\frac{\kappa_s}{\kappa_p} \; \frac{1}{{{N}_{th}
    T_F}}}
    \end{eqnarray}
In order to estimate this restriction let us choose following
parameters \cite{PateraG.2010}: pump wavelength $\lambda_p=0.4
\mu\mbox{m}$, threshold power of continuous generation $P_{th}=50$W,
averaging time (which is equal to the correlation time) $T_F=10$fs
and $\kappa_p = 10 \kappa_s$. As a result one gets
    \begin{eqnarray}
    && \mu_0-1 \gg 10^{-3}.
    \end{eqnarray}

\newpage

    \begin{figure}
    \centering
    \includegraphics[width=10cm]{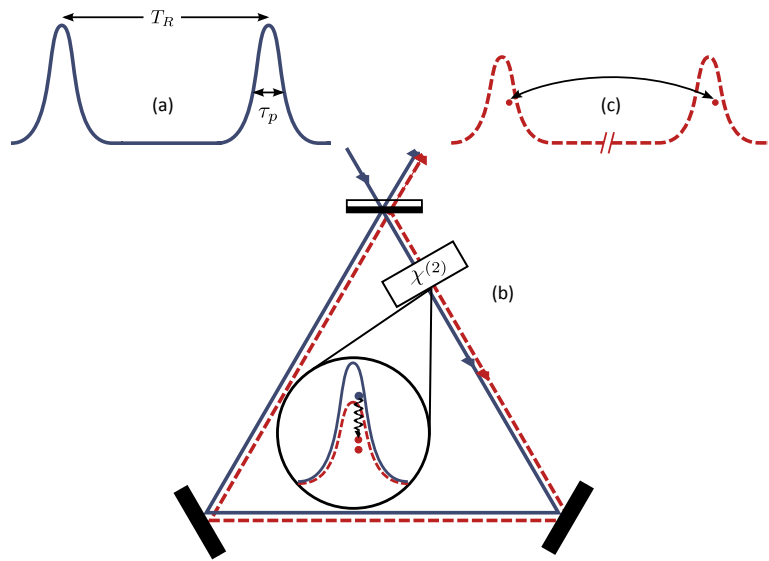}
    \caption{Synchronously pumped optical parametric oscillator (SPOPO):
    (a) pulsed pumping field; (b) parametric down-conversion of pump photon
    in a pair of signal photons inside a nonlinear crystal;
    (c) establishing of quantum correlations between
    pulses of output signal field. Solid line - pump field;
    dashed line - signal field.}
    \label{fig:SPOPO}
    \end{figure}
    \begin{figure}
    \centering
    \includegraphics[width=10cm]{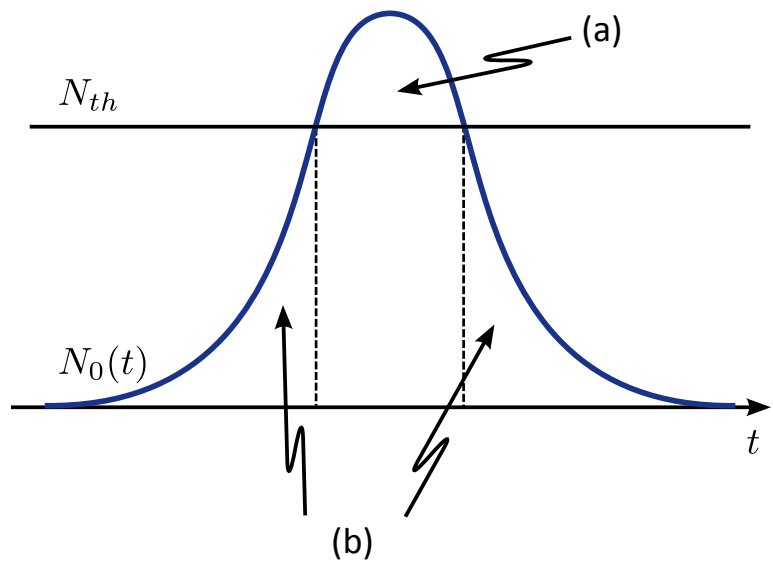}
    \caption{
    Two regimes of oscillation within an individual pump pulse of
    SPOPO:
    (a) above threshold regime; (b) below threshold regime.
    $N_{th}$ - threshold power of continuous generation;
    $N_0(t)$ - instantaneous power of pulsed pumping.}
    \label{fig:threshold}
    \end{figure}
    \begin{figure}
    \centering
    \includegraphics[width=10cm]{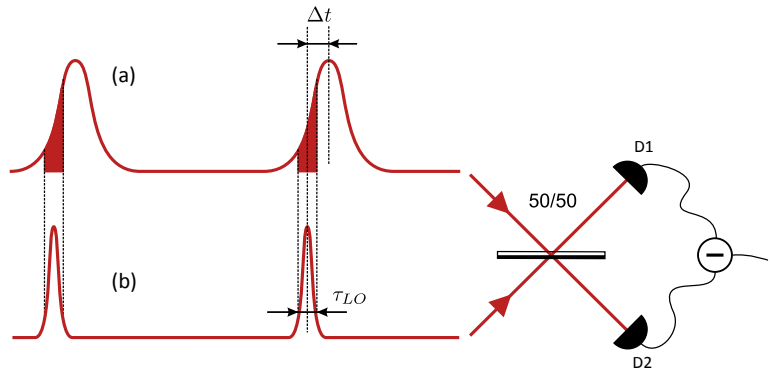}
    \caption{Balanced homodyne detection of a pulsed field:
    (a) analyzed field;
    (b) local oscillator.
    The parameters that can be changed in the setup
    are the duration of the local oscillator pulses $\tau_{LO}$ and their
    delay $\Delta t$ relative to the signal pulses.}\label{fig:BHD}
    \end{figure}
    \begin{figure}
    \centering
    \includegraphics[width=10cm]{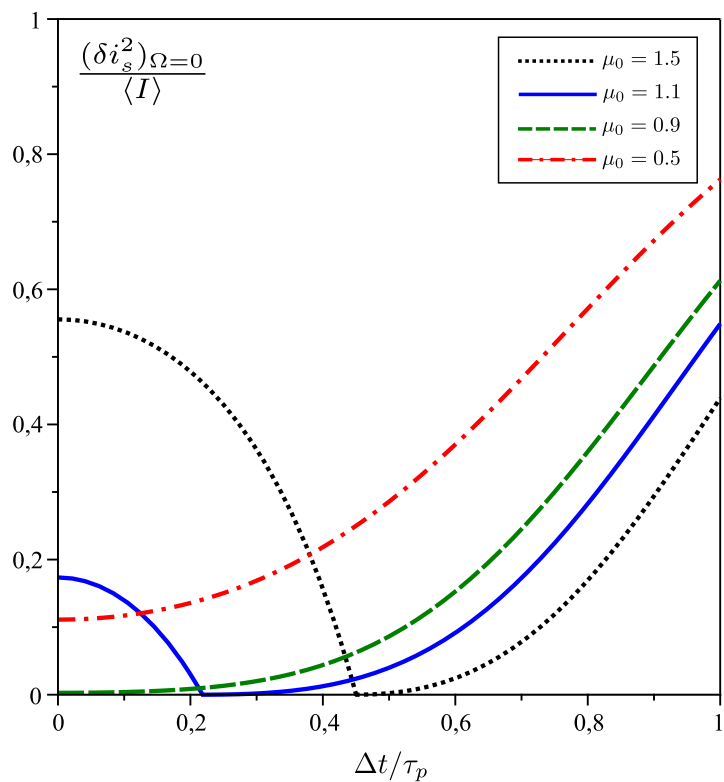}
    \caption{
    Shot noise reduction of difference photocurrent at zero frequency in
    dependence of delay $\Delta t$ of short local oscillator pulses
    ($\tau_{LO}\ll \tau_p$) relative to signal ones. Pump pulses are
    Gaussian with different values of peak power $\mu_0$.
    } \label{fig:mode_sqz}
    \end{figure}

\end{document}